# Hybrid-space reconstruction with add-on distortion correction for simultaneous multi-slab diffusion MRI

Jieying Zhang[1], Simin Liu[1], Erpeng Dai[2], Xin Shao[1], Ziyu Li[3], Karla L. Miller[3], Wenchuan Wu[3], and Hua Guo[1]*

[1] Center for Biomedical Imaging Research, Department of Biomedical Engineering, School of Medicine, Tsinghua University, Beijing, China;
[2] Department of Radiology, Stanford University, Stanford, CA, United States;
[3] Wellcome Centre for Integrative Neuroimaging, FMRIB, Nuffield Department of Clinical Neurosciences, University of Oxford, Oxford, United Kingdom.

**Running title:** Reconstruction with distortion correction for SMSlab

**Word Count:** ~5500

**\*Correspondence to:**

Hua Guo, PhD

Center for Biomedical Imaging Research

Department of Biomedical Engineering

Tsinghua University, Beijing, China

Phone: +86-10-6279-5886

Email: huaguo@tsinghua.edu.cn

**Abstract (250)**

**Purpose:** This study aims to propose a model-based reconstruction algorithm for simultaneous multi-slab diffusion MRI acquired with blipped-CAIPI gradients (blipped-SMSlab), which can also incorporate distortion correction.

**Methods:** We formulate blipped-SMSlab in a 4D k-space with $k_z$ gradients for the intra-slab slice encoding and $k_m$ (blipped-CAIPI) gradients for the inter-slab encoding. Because $k_z$ and $k_m$ gradients share the same physical axis, the blipped-CAIPI gradients introduce phase interference in the $z$-$k_m$ domain while motion induces phase variations in the $k_z$-$m$ domain. Thus, our previous k-space-based reconstruction would need multiple steps to transform data back and forth between k-space and image space for phase correction. Here we propose a model-based hybrid-space reconstruction algorithm to correct the phase errors simultaneously. Moreover, the proposed algorithm is combined with distortion correction, and jointly reconstructs data acquired with the blip-up/down acquisition to reduce the g-factor penalty.

**Results:** The blipped-CAIPI-induced phase interference is corrected by the hybrid-space reconstruction. Blipped-CAIPI can reduce the g-factor penalty compared to the non-blipped acquisition in the basic reconstruction. Additionally, the joint reconstruction simultaneously corrects the image distortions and improves the 1/g-factors by around 50%. Furthermore, through the joint reconstruction, SMSlab acquisitions without the blipped-CAIPI gradients also show comparable correction performance with blipped-SMSlab.

**Conclusion:** The proposed model-based hybrid-space reconstruction can reconstruct blipped-SMSlab diffusion MRI successfully. Its extension to a joint reconstruction of the blip-up/down acquisition can correct EPI distortions and further reduce the g-factor penalty compared with the separate reconstruction.

**Keywords:** 3D diffusion MRI, 4D k-space, simultaneous multi-slab EPI, blipped-CAIPI, distortion correction

# 1. Introduction

Diffusion-weighted MRI (dwMRI) is an essential tool for non-invasive examinations of neuroanatomy by probing the diffusive motion of water molecules (1-3). For clinical examinations, high-resolution dwMRI can detect small lesions invisible in low-resolution dwMRI (4). Improving the resolution of dwMRI can also help differentiate small fibers with complex architecture in the white matter (5-7) and characterize layers in the gray matter (8). Spatial resolution is improved at the cost of long acquisition time or reduced signal-to-noise ratio (SNR). Therefore, for high-resolution dwMRI with multiple diffusion directions, improving its SNR efficiency (9) is crucial.

A series of 2D/3D hybrid techniques have been proposed to reduce the TR and thereby improve the SNR efficiency, including 2D simultaneous multi-slice (SMS) (10-18), 3D multi-slab (9,19-27), and simultaneous multi-slab (SMSlab) (28-31) techniques. 2D SMS excites multiple slices simultaneously, encodes within each slice using conventional 2D k-space acquisitions, and separates slices in reconstruction by leveraging the coil sensitivities in the slice direction. While traditional in-plane parallel imaging incurs a stiff SNR penalty because fewer k-space samples are acquired compared to unaccelerated images (32,33), SMS acceleration has minimal SNR reduction because the same number of k-space samples are acquired but from multiple slices. SMSlab is an extension of 2D SMS that excites multiple thick slabs simultaneously which are encoded using conventional 3D k-space techniques. Like 2D SMS, SMSlab increases the spatial coverage along the slice direction without prolonging the TR (28-31) to achieve optimal SNR efficiency, which is about 1.2~2 s for white matter and gray matter at 3T (34).

SMS imaging can be combined with blipped-controlled aliasing in parallel imaging (blipped-CAIPI) to better utilize the spatial variation of coil sensitivities and reduce the g-factor penalty (12). Such an acquisition can be described by a 3D k-space framework ($k_x$-$k_y$-$k_m$) with $k_m$ representing multi-band encoding in the slice direction (35). Blipped-CAIPI has been recently integrated with SMSlab, which is named blipped-SMSlab, to control the aliasing pattern and reduce the g-factor penalty (36). Accordingly, a 4D k-space framework ($k_x$-$k_y$-$k_z$-$k_m$) is also formulated to model the signal encoding (36). Both $k_z$ (intra-slab slice encoding) and $k_m$ (multi-band or inter-slab encoding) gradients are applied along the slice direction but are conceived of as having different purposes, with $k_z$ defining slices within a given slab and $k_m$ improving the separation of slabs by inducing apparent in-plane shifts. Therefore, modeling $k_z$ and $k_m$ encoding as two conceptual axes rather than as one axis can help to explain the encoding. However, in blipped-SMSlab, $k_z$ and $k_m$ encoding gradients interact with each other and introduce extra phase interference because of the gap between the simultaneously excited slabs (36,37). A correction method has been proposed to solve the problem in a GRAPPA-like reconstruction framework (36), but data have to be transformed between k-space and image space because the phase interference and the motion-induced phase are in different spaces. In this study, the first target is to formulate the phase interference introduced by the blipped-CAIPI gradients and the motion-induced phase in a forward model and reconstruct diffusion images via the model-based method.

For 3D dwMRI, EPI-acquired images also suffer from geometric distortions introduced by $B_0$ field inhomogeneities and eddy currents. The conventional post-processing methods, using a measured $B_0$ map (38-40) or two sets of EPI images with reversed phase encoding (PE) polarities (blip-up/down) (41-44) can be applied for distortion correction. However, reconstructing the blip-up and blip-down images separately does not utilize the data adequately in distortion correction. Thus, a joint reconstruction of the blip-up/down data has been reported to correct image distortions and reduce the g-factors simultaneously (45,46). Similarly, the phase accrual that introduces geometric distortions can be incorporated into the degradation model proposed in this study. Therefore, the second target of this study is to take one step further and jointly reconstruct the blip-up/down data of blipped-SMSlab.

Taken together, we aim to develop an efficient model-based reconstruction method for blipped-SMSlab that avoids having to iterate between k-space and image space. This approach has the added benefit of being able to elegantly incorporate distortion correction and reduce g-factor penalty when blip-up/down data are acquired.

## 2. Theory

*2.1. The 4D k-space framework for blipped-SMSlab*

A 4D k-space framework ($k_x$-$k_y$-$k_z$-$k_m$) is formulated to model the signal encoding of blipped-SMSlab (36). Figure 1A shows an example in which two slabs (Slab A and B) are simultaneously excited. Although the intra-slab slices and the simultaneously excited slabs are both encoded by the gradients along the slice direction, they can be conceptually separated into two logical dimensions in k-space, namely $k_z$ (intra-slab slice encoding) and $k_m$ (multi-band or inter-slab encoding). For $k_z$ encoding, the FOV is defined as the slab thickness $FOV_z$ (including slices for over-sampling). For multi-band ($k_m$) encoding, the FOV is defined as $FOV_m = R_{mb} \cdot z_{\text{gap}}$, where $R_{mb}$ is the multi-band factor and $z_{\text{gap}}$ is the distance between the center of two neighboring simultaneously excited slabs; that is, $FOV_m$ reflects the full extent of excited tissue. As for the in-plane encoding, conventional frequency encoding ($k_x$) and phase encoding ($k_y$) are used. The corresponding representations of $k_x$, $k_y$, $k_z$, and $k_m$ in image space are x, y, z, and m, respectively.

Although $k_z$ and $k_m$ encoding are separate in the trajectory (Figure 1C), they come from the same direction gradient, which are represented by $G_z$ and $G_m$, respectively, in Figure 1D. The blipped-CAIPI ($G_m$) gradients are applied during the EPI readout. When SMSlab EPI is combined with blipped-CAIPI gradients, the trajectory of one EPI echo train is similar to that of 2D SMS EPI with blipped-CAIPI gradients, which is not retained in a single 2D $k_x$-$k_y$ plane. Instead, the $k_y$ lines are shifted along the $k_m$ dimension (blue trajectory in Figure 1C), which introduces slab shifting in image space and enables better use of the spatial variation of coil sensitivities along the slab direction. As for the $k_z$ encoding, it encodes the slices within each slab through a table of $G_z$ gradients before the EPI readout. By introducing the $k_m$ dimension, the trajectory can be presented as acquiring a $k_z$ plane in one shot, despite the fact that the blipped-CAIPI gradients are varying the gradient strength during the acquisition.

*2.2. Phase interference introduced by the $k_z$ and $k_m$ gradients*

According to our previous study (36), the $k_z$ and $k_m$ gradients interact with each other because they share the same physical axis. They bring extra phase interference to the signals along $k_z$ and $k_m$ dimensions due to the gap between the simultaneously excited slabs, which includes two parts (Figure 2 in (36)).

First, due to the gap between the centers of the simultaneously excited slabs ($z_{\text{gap}}$), an extra phase offset $\varphi_{\text{gap}}$ is introduced to the slab deviating from the isocenter (Slab B in Figure 1A) in the $k_z$-m domain that is linear along $k_z$. If without $\varphi_{\text{gap}}$ correction, a slice shift occurs along z dimension within slab B (Figure 5 in (36)):

$$\varphi_{\text{gap}} = 2\pi \frac{z_{\text{gap}}}{FOV_z} n_{kz} n_m. \qquad (1)$$

$FOV_z$ is the slab thickness (including over-sampling). $n_{kz}$ is the index of $k_z$ encoding along the intra-slab slice dimension in k-space, and $n_m$ is the slab index in image space. Since each $k_z$ plane is acquired in one shot, $\varphi_{\text{gap}}$ is constant for a given excitation. RF phase modulation can therefore be used to compensate for it by introducing $-\varphi_{\text{gap}}$ to the multi-band RF pulses (36).

Second, the blipped-CAIPI gradients applied during the EPI readout to aid in separating slabs in the conceptual $k_m$ dimension also introduce shifts along the $k_z$ dimension. This means the acquired samples deviate from the nominal k-space locations (Figure 1B). The $k_z$ shifts are equivalent to additional linear phase modulation along z (the intra-slab slice dimension), which we refer to as the ramp phase (36).

$$\varphi_{\text{ramp}} = 2\pi \frac{\Delta z}{FOV_m} n_{km} n_z. \qquad (2)$$

$\Delta z$ is slice thickness. $FOV_m$ is the defined FOV in the multi-band dimension. $n_{km}$ is the index of $k_m$ encoding along the multi-band dimension in k-space, and $n_z$ is the intra-slab slice index in image space.

Because $\varphi_{\text{ramp}}$ is in the z-$k_m$ domain, it needs to be accounted for in the reconstruction. In theory, this is a straightforward correction based on a priori knowledge of the applied gradients. However, a complicating factor is motion-induced phase variations that vary unpredictably from shot to shot and are therefore in the $k_z$-m domain. In our previous study, a GRAPPA-like reconstruction is used that takes several steps to correct $\varphi_{\text{ramp}}$ and the motion-induced phase. Because they are located in different domains, meaning that data must be transformed back and forth between k-space and image space (36). In the present study, the correction for both phases is achieved by integrating $\varphi_{\text{ramp}}$ and the motion-induced phase into a forward model, which is illustrated in the following sections.

## 3. Methods

### 3.1. Blipped-SMSlab sequence with a blip-up/down acquisition

The blipped-SMSlab sequence is shown in Figure 1D (also Figure 3 of (36)). The multi-band RF pulses used for excitation and refocusing are the summation of multiple single-band pulses (one per simultaneously-excited slab). RF phase modulation is used for $\varphi_{\text{gap}}$ correction, as mentioned above. A table of $k_z$ gradients is used before the EPI readout to encode slices within each slab. One EPI plane in $k_x$-$k_y$ is acquired for each excitation, corresponding to a single $k_z$ plane in a basic blipped-SMSlab sequence. During the first echo train, blipped-CAIPI gradients ($G_m$) are applied simultaneously with the $G_y$ blips for multi-band (or inter-slab) encoding.

To work in practice, this method also has to address motion artifacts that are inherent to 3D dwMRI. Navigators are usually acquired in 3D dwMRI (6,9,19,24,31,47). Though a one-shot 3D navigator can record the through-plane motion ideally, it's not commonly used because of the prolonged TE (48). Therefore, a 2D SMS navigator is acquired (30,36) (indicated as echo_2 in Figure 1D) with the assumption that the motion-induced phase does not vary too much along the slice direction if the slab thickness does not exceed 20 mm (9,19,21,22). The simultaneously excited slices of the navigator must be recovered in the reconstruction process. Blipped-CAIPI gradients are not used for the navigator because of higher SNR at lower resolution. The effective echo spacing (ESP) of the navigator stays consistent with that of the first echo train to guarantee the same distortions. It is achieved by adjusting the matrix size in the readout direction or the undersampling factor in the PE direction. The PE polarity of the navigator is the same as the first echo train.

Furthermore, we propose to combine blipped-SMSlab with the blip-up/down acquisition for distortion correction. The trajectory and the sequence diagram are shown in Figures 1C and 1D. The blip-up/down acquisition refers to alternating the direction of $k_y$ traversal (by negating the polarities of the $k_y$ blip gradients in the EPI readouts) across different shots. For each diffusion-weighted image, two acquisitions with blip-up/down acquisition are acquired for each $k_z$ plane. To reduce the inter-shot motion between the blip-up/down pairs, they are acquired inside the $k_z$ loop in sequence programming. To minimize the cross-talk effects between neighboring slabs, the slabs are excited in an interleaved way, which is the innermost loop.

*3.2. Reconstruction for blipped-SMSlab in hybrid space*

A core concept is which k-space axes can simply be Fourier transformed and which require additional reconstruction steps. Among the 4 dimensions of the encoding, $k_y$ and $k_m$ dimensions are undersampled while the motion-induced phase differs across $k_z$ steps. Therefore, inverse Fast Fourier Transform (iFFT) is first applied along $k_x$ to transform the k-space signals into the hybrid space (x-$k_y$-$k_z$-$k_m$), which enables a separate reconstruction for each x index (49). The k-space signals in this hybrid space are represented by d. Assuming that there are no motion-induce phase variations, the 4D representation of blipped-SMSlab signals can be formulated as

$$(F_u \circ \Phi)SI = d, \qquad (3)$$

where $F_u$ is a 3D FFT operator on y, z, and m dimensions with undersampling along $k_y$ and $k_m$ dimensions. The correction of $\varphi_{\text{ramp}}$ is achieved by multiplying the ramp phase ($\Phi(n_z, n_{km}) = \varphi_{\text{ramp}}$) with the FFT operator. $\circ$ means element-wise product. S represents coil sensitivities. $I$ is the fully sampled image.

For b=0 images, $\varphi_{\text{ramp}}$ can be directly removed from the k-space signals after 1D iFFT along $k_z$ (36). For diffusion-weighted images, 1D iFFT along $k_z$ cannot be applied directly because of motion-induced phase variations, which should also be taken into consideration. Then the forward model becomes

$$(F_u \circ \Phi)(S \circ P)I = d, \qquad (4)$$

and P represents motion-induced phase variations.

Based on the forward model, a reconstruction algorithm called REconstruction with phAse Correction in a Hybrid space (REACH) is proposed, and the flowchart is shown in Figure 2. In step 1, the motion-induced phase maps extracted from the reconstructed navigators are smoothed by a Gaussian filter with a width of 10. In step 2, the diffusion-weighted data are transformed to the x-$k_y$-$k_z$-$k_m$ space (with $k_y$ and $k_m$ undersampled) and reconstructed using the following cost function via Landweber iteration (50). The reconstructed images are represented by $I_{REACH}$.

$$I_{REACH} = \underset{I}{\mathrm{argmin}} \|d - (F_u \circ \Phi)(S \circ P)I\|_2^2. \qquad (5)$$

Specifically, the navigator reconstruction process is shown in Supporting Information Figure S1. For the navigator of each shot, conventional 2D SENSE reconstruction (33) is applied in the multi-band dimension to separate the overlapped slices. If undersampling in the PE direction ($k_y$) is also used, 3D SENSE is used instead. The complex conjugates of the phase differences between diffusion-weighted navigators and b=0 navigators are used as P.

### 3.3. Reconstruction with distortion correction

The REACH algorithm is able to incorporate distortion correction when blip-up/down data are acquired, which can reduce the g-factor penalty simultaneously (46). The extension of REACH for distortion correction is called Distortion-Corrected REACH (DC-REACH). Following the previous work (45,46), the blipped-SMSlab signals with field-inhomogeneity-induced phase accrual can be formulated as

$$(F_u \circ \Phi \circ \Theta)(S \circ P)I = d, \qquad (6)$$

where $\Theta$ is the field-inhomogeneity-induced phase accrual along $k_y$ dimension which causes image distortions. $\Theta$ is formulated as $\Theta(\Delta f, n_{ky}) = e^{i \cdot \Delta f \cdot (ESP_{eff} \cdot n_{ky})}$, where $n_{ky}$ is the $k_y$ index, and $ESP_{eff}$ is the effective echo spacing. For b=0 data, $\Delta f$ represents the off-resonance field map. For diffusion-weighted data, $\Delta f$ contains both the static off-resonance field and the transient eddy currents field. To jointly reconstruct the blip-up and blip-down data, the background phase difference D caused by the reversed PE polarities needs to be corrected (46):

$$(F_u \circ \Phi \circ \Theta)(S \circ D \circ P)I = d. \qquad (7)$$

Here d is the concatenated data of the blip-up and blip-down signals in the x-$k_y$-$k_z$-$k_m$ space.

The process of DC-REACH is shown in Figure 3. First, the blip-up and blip-down data are separately reconstructed using REACH (steps 1-2 in Figure 2) without distortion correction.

Then, in step 3, the image amplitudes of the blip-up/down pairs are used to estimate the field maps $\Delta f$ using *topup* (41) from the FMRIB Software Library (FSL) (51). The operator $\Theta$ is generated from these field maps, which capture both the static off-resonance field and the eddy currents induced by the diffusion gradients. In step 4, the background phase difference $D$ is generated by subtracting the blip-up phase map from the blip-down phase map, and the motion-induced phase $P$ is extracted from the navigators. The effect of $D$ correction is shown in Supporting Information Figure S3. It is worth noting that before the phase extraction, we apply distortion correction and smoothing to both the images reconstructed in step 2 and the navigators (Supporting Information Figure S2), to reduce the impact of geometric mismatch between the phase maps of the blip-up and blip-down acquisitions on the results of DC-REACH (Supporting Information Figure S4). The real and the imaginary part of the complex images are distortion-corrected separately via the field mapping method (38) using $\Delta f$ generated in step 3. In step 5, DC-REACH minimizes the cost function in Eq. 8 using Landweber iteration (50).

$$I_{DC} = \underset{I}{\operatorname{argmin}} \| d - (F_u \circ \Phi \circ \Theta)(S \circ D \circ P)I \|_2^2. \tag{8}$$

In the implementation, the ramp phase $\Phi$ and the distortion operator $\Theta$ are combined with the FFT operator $F_u$, and the motion-induced phase $P$ and the background phase difference $D$ are combined with coil sensitivities $S$ before the iteration.

*3.4. Experiments*

To evaluate the proposed algorithms, four in-vivo experiments on three healthy subjects and one phantom experiment were conducted in this study. The detailed acquisition parameters are listed in Table 1. All images were acquired on a 3T scanner (Ingenia CX, Philips Healthcare, Best, The Netherlands) using a 32-channel head coil. All human studies were performed under the institutional review board approval from our institution, and written consent was given before the study. All of the SMSlab, blipped-SMSlab, and single-band multi-slab (3D multi-slab imaging without multi-band encoding) sequences were acquired with blip-up/down acquisitions for distortion correction.

In the first in-vivo experiment, images were acquired on the first subject to demonstrate the feasibility of the proposed reconstruction frameworks (REACH and DC-REACH) and evaluate the effects of blipped-CAIPI gradients. A blipped-SMSlab sequence was applied with the blip-up/down acquisition. The resolution was 1.5-mm isotropic and the effective ESP was 0.377 ms. The $k_y$ acceleration factor for each shot ($R_y$, marked in Figure 1C) was 2. Nine slabs with MB=2 and 8 slices for each (including 2 slices for over-sampling) were acquired. After reconstruction, the over-sampled slices were discarded before concatenating all the slabs. TR=1900 ms and TE=70 ms were used. Three orthogonal diffusion encoding directions were acquired with b=1000 s/mm$^2$. Then, a SMSlab sequence without blipped-CAIPI was acquired with the blip-up/down acquisition to be compared with the blipped-SMSlab sequence. For the acquisition without blipped-CAIPI gradients (Supporting Information Figures S6B and E, the simultaneously excited slabs are stitched together along the slice direction to provide a 3D dataset (28,30). For the blip-up/down acquisition (without the blipped-CAIPI gradients), a CAIPI sampling pattern is achieved by shifting the blip-down shots along $k_y$ and $k_z$. The other acquisition parameters were the same as the blipped-SMSlab sequence. Data acquired without blipped-CAIPI were reconstructed by the 3D versions of REACH and DC-REACH without ramp phase correction (52). The details of SMSlab without blipped-CAIPI are introduced in the Supporting Information. The g-factors of both sequences were calculated via a Monte-Carlo-based method with 128 repetitions (53).

The second experiment was conducted on the second subject to evaluate the residual errors of the REACH and DC-REACH reconstruction. The same SMSlab sequences (with and without blipped-CAIPI) as the first experiment were applied with 1.5-mm isotropic resolution. A fully sampled SMSlab sequence (28,30) was applied to acquire the reference image of the REACH reconstruction. It used an MB=2 excitation, but the simultaneously excited slabs were fully sampled along the slice direction. Two interleaved shots were used for each $k_z$ plane to match the distortions of the undersampled sequences. The fully sampled acquisition was repeated twice with reversed PE polarities. Then, the reference image was distortion-corrected using *topup* and *applytopup* (41) from the FSL (51) and used as the reference of the DC-REACH reconstruction. Only b=0 images were acquired as they are more robust to motion and more suitable for the evaluation compared with diffusion-weighted images.

The third experiment was conducted on an ACR phantom to evaluate the performance of DC-REACH for distortion correction. The same blipped-SMSlab sequence as the first experiment was used. In the first and third experiments, T2W images (distortion-free reference) were acquired using a 3D turbo spin echo (TSE) sequence with 1-mm isotropic resolution, TR=3000 ms, and TE=280 ms.

In the fourth experiment on the third subject, a DTI dataset with 1.3-mm isotropic resolution was acquired using blipped-SMSlab to evaluate the quantitative DTI metrics. $R_y$=3 along ky was used to reduce the effective ESP to 0.381 ms. Seven slabs with MB=2 and 10 slices for each (including 2 slices for over-sampling) were acquired. TR=1800 ms and TE=70 ms were used. Two b=0 s/mm$^2$ images and 40 b=1000 s/mm$^2$ images were acquired. Fluid attenuated inversion recovery (FLAIR) images were acquired using 3D TSE with 1-mm isotropic resolution, inversion time=1650 ms, TR=4500 ms, and TE=320 ms.

The fifth experiment was conducted on the first subject to show that DC-REACH can be adapted for single-band multi-slab imaging. The resolution, b value, and effective ESP were the same as those in the first experiment. $R_y$=2, TR=3800 ms, and TE=62 ms were used. Eighteen slabs with MB=1 and 8 slices for each (including 2 slices for over-sampling) were acquired. For single-band multi-slab imaging, two excitations were used in each $k_z$ with reversed PE polarities, as shown in Supporting Information Figure S6C. The forward model and the results are shown in the Supporting Information. As the distortion-free reference, 3D T2W TSE images were also acquired.

For all experiments, volumetric $B_0$ shimming was used. Coil sensitivities were measured using a fully sampled distortion-free 2D gradient-echo sequence with a matrix size of 64. Its slice thickness was equal to the slab thickness of the SMSlab scan. The sensitivity maps S were estimated using ESPIRiT (54). For both REACH and DC-REACH, 40 iterations with a step size of 0.3 were used in the Landweber iteration, and a zero matrix was used as the initial guess of the image. The reconstruction algorithms were implemented using MATLAB on a server with an AMD Ryzen 9 3900X CPU and 128G memory. A slab profile scan was conducted to estimate the initial slab profiles for slab boundary artifact correction in each experiment. It uses the same sequence as the single-band multi-slab or SMSlab scan but without a blip-up/down acquisition and diffusion encoding. An over-sampling factor of 100% was used along the intra-slab slice ($k_z$) dimension to avoid aliasing.

*3.5. Pre- and Post-processing*

The acquired data were divided into 2 datasets, blip-up and blip-down, before the pre-processing. First, the correction for the off-isocenter phase error was done as reported in the previous study (36). N/2 ghost artifacts were corrected using an SVD-based method (55).

Moreover, slab boundary artifacts must be corrected in single-band multi-slab and SMSlab imaging, which are mainly introduced by the truncated RF pulses (21,22,28,56,57). Convolutional-neural-network-enabled inversion for slab profile encoding (CPEN) is a model-based deep learning method, which outperforms the other algorithms on SMSlab imaging (56). It was used for slab boundary artifact correction in this study. The slab profiles were estimated using the slab profile scan described in the previous section. Note that for the images by DC-REACH, the slab profile images were distortion-corrected by the field mapping method (38) to match the geometry of the reconstructed images.

Finally, co-registration of different diffusion directions (58) and diffusion tensor imaging (DTI) fitting (59) were applied using the *eddy* and *dtifit* commands from FSL (51), respectively.

## 4. Results

The blipped-SMSlab images with 1.5-mm isotropic resolution (from the first experiment) are used to show the effect of $\varphi_{\text{ramp}}$ correction (Figure 4 and Supporting Information Figure S5). Images are rescaled for comparison. Error maps are generated by subtracting the magnitude of the images with $\varphi_{\text{ramp}}$ correction from that of the images without $\varphi_{\text{ramp}}$ correction, and then normalizing the errors by the maximum amplitude of the images with $\varphi_{\text{ramp}}$ correction. Therefore, positive values in the error maps represent extra artifacts introduced to the images if $\varphi_{\text{ramp}}$ correction is not done. For the diffusion-weighted images, without $\varphi_{\text{ramp}}$ correction, residual ghost-like aliasing exists in the images (yellow arrows in

Figure 4). For the b=0 s/mm$^2$ images, $\varphi_{\text{ramp}}$ correction is completed in the pre-processing. Ghost-like aliasing is also found in the error maps (yellow arrows in Supporting Information Figure S5).

Figure 5 shows the reconstructed blipped-SMSlab images with 1.5-mm isotropic resolution from the first experiment. Results from both REACH and DC-REACH are shown. T2W images serve as the distortion-free reference. In DC-REACH, the distortions, especially around the forehead, are corrected, and the geometry is close to the T2W images. The corresponding 1/g-factor maps are shown at the bottom. For the b=0 s/mm$^2$ images, the mean values of 1/g-factor of REACH and DC-REACH are 0.42 and 0.65, respectively. For the b=1000 s/mm$^2$ images, the mean 1/g-factor values are 0.44 and 0.67, respectively. Compared with the separate REACH, DC-REACH utilizes the acquired data more effectively and distinctly increases the 1/g-factor, indicating that a higher SNR is obtained via the joint reconstruction of the blip-up and blip-down data. The computation time of 1 diffusion direction is 52 mins and 93 mins for REACH and DC-REACH, respectively.

The effect of blipped-CAIPI on REACH and DC-REACH is evaluated via two comparisons between the SMSlab acquisitions with and without blipped-CAIPI (Figure 6 and Supporting Information Figure S7). The corresponding sampling trajectories are shown in Supporting Information Figures S6A and B. The images from the first experiment and the corresponding 1/g-factor maps are shown in Figure 6. Blipped-CAIPI gradients increase the 1/g-factor in the separate REACH for the blip-up/down data, compared with SMSlab without blipped-CAIPI (left panel in Figure 6). The mean value of 1/g-factor improves from 0.39 to 0.45. However, for DC-REACH, no distinct difference in the histograms of 1/g-factor is found between the images with and without blipped-CAIPI (right panel in Figure 6). For blipped-SMSlab, the mean value of 1/g-factor improves by 47% in DC-REACH compared with REACH. For SMSlab without blipped-CAIPI, the improvement is around 72% in this experiment.

Data from the second experiment are shown in Figure S7. Compared with the reference image acquired by the fully sampled SMSlab sequence, the normalized root mean squared errors (NRMSEs) are 17.3% and 16.1% for undersampled SMSlab without and with blipped-CAIPI, respectively, in the REACH-reconstructed images without distortion correction (Figure

S7, upper left). The field maps of both SMSlab sequences are slightly different from the map estimated from the reference image (Figure S7, bottom left). As for the results of DC-REACH, the NRMSEs are 15.0% and 15.2% for SMSlab without and with blipped-CAIPI, respectively (Figure S7, upper right). This will be further explained in the discussion section later.

The phantom images from the third experiment are used to evaluate the performance of distortion correction (Figure 7). The T2W images serve as the distortion-free reference. The edges of the grids in the T2W images are extracted and superimposed onto the reconstructed EPI images. The geometry of the images reconstructed with DC-REACH agrees with the T2W images well. The distortions are thus well corrected using the proposed method. The off-resonance frequencies range from -45.8 Hz to 95.1 Hz, and the standard deviation is 36.9 Hz.

Figure 8 compares the in-vivo images reconstructed with DC-REACH with the T2W images in three orthogonal planes from the first experiment. In most regions, the distortions are well-corrected and the geometry is close to the T2W images. However, DC-REACH may not fully recover the slight signal loss at the air-tissue interface (yellow arrow), which results from the severe $B_0$ inhomogeneities. Although the standard deviation of the field map is only 39.1 Hz, the off-resonance frequencies range from -246.2 Hz to 123.3 Hz.

The results of the DTI dataset from the fourth experiment are shown in Figure 9. The b=1000 s/mm$^2$ images, the mean diffusion-weighted images, and the colored fractional-anisotropy (FA) maps are shown in three orthogonal planes. Most of the distortions are corrected. For example, the yellow arrows point out the distortions in the frontal lobe. Around the brainstem and the cerebellum, where severe distortions exist, the slight signal loss remains uncorrected (white arrow).

## 5. Discussion

This study proposed a reconstruction algorithm named REACH for blipped-SMSlab. It reconstructs 3D SMSlab EPI data acquired with blipped-CAIPI in the hybrid space and corrects the phase interference during the reconstruction. Furthermore, it is extended to jointly reconstruct the blip-up and blip-down data for distortion correction, which is called DC-REACH. It can reduce the g-factor penalty simultaneously compared to the conventional topup-based correction. DC-REACH can also be adapted for SMSlab EPI without blipped-CAIPI and single-band multi-slab EPI (Supporting Information Figure S8).

Blipped-SMSlab is a combination of SMSlab EPI and blipped-CAIPI gradients. Because the $k_z$ gradients and the $k_m$ gradients share the same physical axis, extra phase interference is introduced to these two dimensions. The $k_m$ gradients impose an extra group of linear phases $\varphi_{\text{ramp}}$ on the slices within each slab. The previous study showed how to correct $\varphi_{\text{ramp}}$ in a GRAPPA-like reconstruction method (36). Because $\varphi_{\text{ramp}}$ is in the z-$k_m$ domain, while the motion-induced phase variation is in image space, they need to be corrected separately in the k-space-based reconstruction process. The forward model proposed in this study contains an operator for FFT, which makes it possible to model the phase errors both in the k-space and image space. Therefore, $\varphi_{\text{ramp}}$, together with the motion-induced phase variations measured by navigators, can be corrected in a one-step model-based reconstruction.

Moreover, REACH enables the combination with the other model-based methods. In this study, REACH is integrated with distortion correction (DC-REACH). Image distortions are the results of the field-inhomogeneity-induced phase accrual in k-space (46). The phase accrual can also be included in the proposed forward model, which enables a joint reconstruction with distortion correction for blip-up and blip-down data. Moreover, modeling the phase accrual in k-space can describe the continuity of geometric distortions (or pixel shifts) in EPI images, which is discretized if an image-based (SENSE-like) reconstruction method is used (60).

DC-REACH can also reduce the g-factors, compared with the separate REACH reconstruction of the blip-up or blip-down dataset (Figure 5). The effect of blipped-CAIPI is also evaluated (Figure 6 and Supporting Information Figure S7). Blipped-CAIPI gradients can reduce the g-factor penalty in REACH (Figure 6, left panel). However, it may not bring extra benefit to DC-REACH compared with SMSlab without blipped-CAIPI (Figure 6, right panel). The results of the reconstruction error evaluation also support the conclusion. In REACH reconstruction, blipped-CAIPI gradients help to reduce the reconstruction error (Figure S7, upper left), but only slight differences are found in the estimated field maps because they are smoothed (Figure S7, bottom left). Blipped-CAIPI does not distinctly reduce the residual errors in DC-REACH reconstruction (Figure S7, upper right). For SMSlab without blipped-CAIPI, a

CAIPI sampling pattern can be achieved in the blip-up/down acquisition via applying shifts along $k_y$ and $k_z$ for the blip-down shots, as illustrated in the Supporting Information (Figure S6). As a result, more gains in 1/g-factor can be achieved via the joint reconstruction of the blip-up and the blip-down data, compared with blipped-SMSlab. Therefore, the application of blipped-CAIPI gradients is recommended in a basic SMSlab acquisition, which can help the separation of simultaneously excited slabs and reduce the g-factor penalty. For SMSlab scans with the blip-up/down acquisition, an acquisition either with or without blipped-CAIPI gradients can be used, because they may have similar performance concerning the g-factor penalty or reconstruction errors in the joint reconstruction of blip-up/down data (DC-REACH).

The 1/g-factor improvement of blipped-CAIPI in REACH reconstruction is limited for the setting used in this study. Because only MB=2 was used, the distances between two simultaneously excited slabs were larger than 70 cm in our experiments, hence the coil sensitivities of the two slabs are very different. The advantage of blipped-CAIPI can not be fully valued. MB>2 with smaller slab gaps will be tried in the future if TE can be shortened by either reducing the RF pulse duration or shortening the echo train by using powerful gradients.

Although the distortions on the phantom are well corrected, the proposed method performs less well in the in-vivo experiments. It inherits the legacy of topup distortion correction capability (41). According to Figures 6 and 8, the maximum off-resonance frequency in the in-vivo experiment is much larger than that in the phantom experiment, which is mainly located in regions around the air-tissue interface, such as the skull base and the acoustic meatus. This limitation can be more obvious when a high image resolution is used. There are two ways to improve the performance of distortion correction. The first one is to reduce $B_0$ inhomogeneity by dynamic $B_0$ shimming, which can reduce the distortions and the g-factors of DC-REACH (45). The second one is to reduce the effective ESP at the cost of SNR by either using a powerful gradient system or increasing the acceleration factors.

This study only serves as a preliminary investigation for REACH and DC-REACH. Higher resolution and higher acceleration factors can be tried in the future. The highest resolution is 1.3-mm isotropic in this study, and the acceleration factor for REACH is limited to MB×$R_y$=3 ×2. As higher resolution leads to longer ESP and thus severer T2* blurring accordingly, a higher in-plane acceleration factor should be used to shorten the echo train length. DC-REACH can be adapted for single-band multi-slab EPI (Figure S8), but further acceleration needs to be explored. To improve the acceleration factors, advanced regularization terms can be combined with the proposed algorithm to retain the SNR. For example, the SPIRiT regularization is proven to be more effective than traditional reconstruction in 2D multi-shot EPI imaging (61). Moreover, a receive coil with more channels would also help.

So far, the computation time is long due to the large matrix size of SMSlab or blipped-SMSlab images. REACH enables a separate reconstruction for each x index. Therefore, parallel computing and iteration algorithms with the ability of fast convergence can be explored in the future to accelerate the computation.

## 6. Conclusion

In this study, we proposed a hybrid-space reconstruction algorithm, REACH, for SMSlab EPI data acquired with blipped-CAIPI gradients (blipped-SMSlab). It is an efficient model-based method that can solve the phase interference introduced by the blipped-CAIPI gradients in one step. This approach can elegantly incorporate distortion correction (DC-REACH) when using the blip-up/down acquisition, which also reduces the g-factor penalty via the joint reconstruction even if blipped-CAIPI is not used. It can also be extended to different single-band multi-slab EPI sampling strategies.

**Acknowledgements**

The authors would like to thank Dr. Ying Chen for helpful discussions about the 4D k-space theory and Dr. Congyu Liao at Stanford University for helpful discussions about the reconstruction method. This work was supported by Beijing Municipal Natural Science Foundation (L192006) and the National Natural Science Foundation of China (61971258).

**Table:**

Table 1:   Acquisition parameters of the EPI sequences

| Exp. | Sequence | Resolution [mm$^3$] | TE / TR [ms] | Partial Fourier | $R_y$ | MB | Slabs | Slices per slab (target +over-sampled) | $b$ values [s/mm$^2$] (directions) | Scan time [min:sec] |
|---|---|---|---|---|---|---|---|---|---|---|
| 1 | SMSlab | 1.5×1.5×1.5 | 70/1900 | 0.630 | 2 | 2 | 9 | 6+2 | 0 (1), 1000(3) | 2:03 |
|   | Blipped-SMSlab | 1.5×1.5×1.5 | 70/1900 | 0.630 | 2 | 2 | 9 | 6+2 | 0 (1), 1000(3) | 2:03 |
| 2 | SMSlab | 1.5×1.5×1.5 | 70/1900 | 0.630 | 2 | 2 | 9 | 6+2 | 0 (1) | 0:31 |
|   | Blipped-SMSlab | 1.5×1.5×1.5 | 70/1900 | 0.630 | 2 | 2 | 9 | 6+2 | 0 (1) | 0:31 |
|   | Fully sampled SMSlab * | 1.5×1.5×1.5 | 70/1900 | 0.630 | 2 | 2 | 9 | 6+2 | 0 (1) | 2:03 |
| 3 | Blipped-SMSlab | 1.5×1.5×1.5 | 70/1900 | 0.630 | 2 | 2 | 9×9 | 6+2 | 0 (1), 1000(3) | 2:03 |
| 4 | Blipped-SMSlab | 1.3×1.3×1.3 | 70/1800 | 0.654 | 3 | 2 | 7 | 8+2 | 0 (2), 1000(40) | 25:12 |
| 5 | Single-band multi-slab | 1.5×1.5×1.5 | 62/3800 | 0.630 | 2 | 1 | 18 | 6+2 | 0 (1), 1000(3) | 4:03 |

$R_y$: acceleration factor along PE direction for each shot or $k_z$ step.

Exp.: experiment.

In-plane FOV=220×220 mm$^2$.

FOV along slice direction = slice resolution × target slices per slab × slabs × MB.

No gaps between the slabs.

All of the SMSlab, blipped-SMSlab, and single-band multi-slab sequences were acquired with the blip-up/down acquisition for distortion correction.

* Fully sampled SMSlab was repeated twice with reversed phase encoding (PE) polarities. For each acquisition, 2 in-plane shots were used for each $k_z$ plane. It used an MB=2 excitation, but both of the simultaneously excited slabs were fully sampled. No blipped-CAIPI gradients were used.

**Figures:**

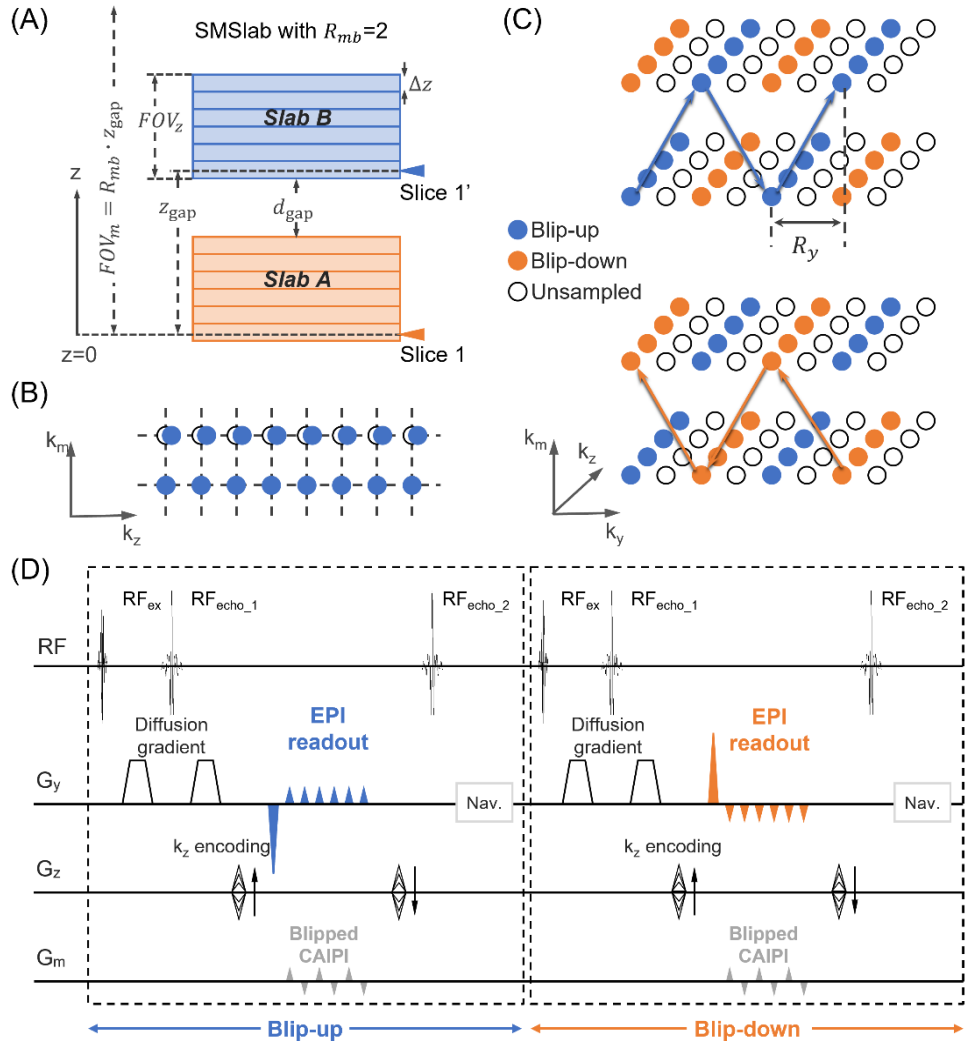

Figure 1: SMSlab EPI with blipped-CAIPI (blipped-SMSlab) combined with the blip-up/down acquisition. (A) shows the target FOV of SMSlab imaging with MB=2, with different colors representing the two simultaneously excited slabs. The grids within each slab represent the intra-slab slices. The isocenter is marked as z=0. (B) shows the deviation along $k_z$ dimension due to the blipped-CAIPI gradients. The dashed lines represent the nominal k-space locations. (C) and (D) show the trajectory and the sequence diagram, respectively. Note that the $G_z$ gradients and the $G_m$ gradients share the same physical axis. Some gradients are omitted in (D) for simplification, including the slice-selection gradients, the $G_x$ gradients, and so on. The diffusion gradients are shown in the $G_y$ direction as an example.

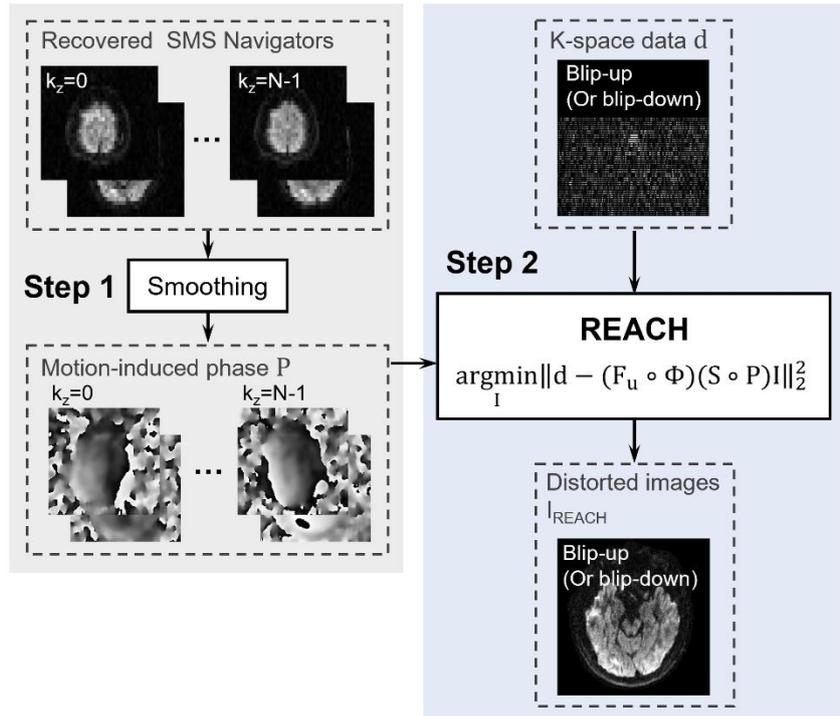

Figure 2: The flowchart of the hybrid-space reconstruction (REACH) for blipped-SMSlab. The navigator reconstruction process is shown in Supporting Information Figure S1.

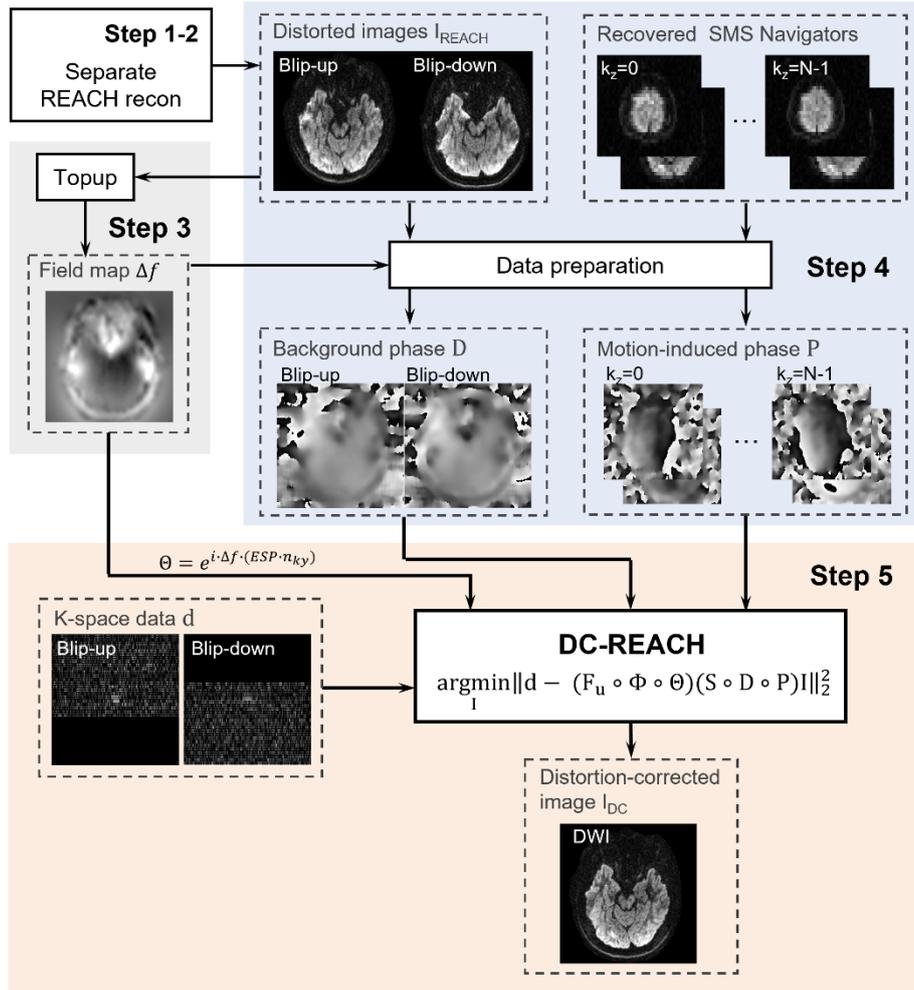

Figure 3: The flowchart of distortion-corrected hybrid-space reconstruction for blipped-SMSlab (DC-REACH) with the blip-up/down acquisition. The processing of the phase maps (step 4) is shown in Supporting Information Figure S2.

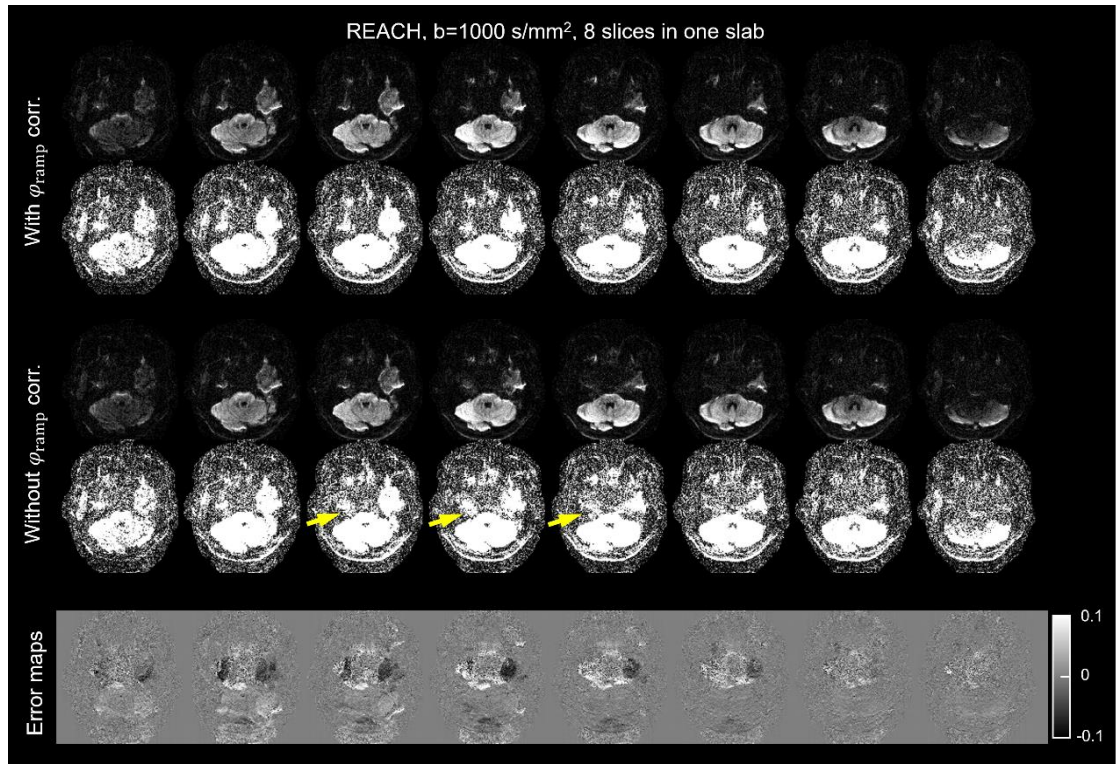

Figure 4: Comparison between the blipped-SMSlab images reconstructed with and without $\varphi_{\text{ramp}}$ correction (corr.). The diffusion-weighted images with 1.5-mm isotropic resolution and b=1000 s/mm$^2$ are shown. Each image is rescaled by a factor of 8. The ghost-like aliasing is pointed out by the yellow arrows. The error maps of the images without $\varphi_{\text{ramp}}$ correction with respect to the images with $\varphi_{\text{ramp}}$ correction are normalized by the maximum amplitude of the images with $\varphi_{\text{ramp}}$ correction.

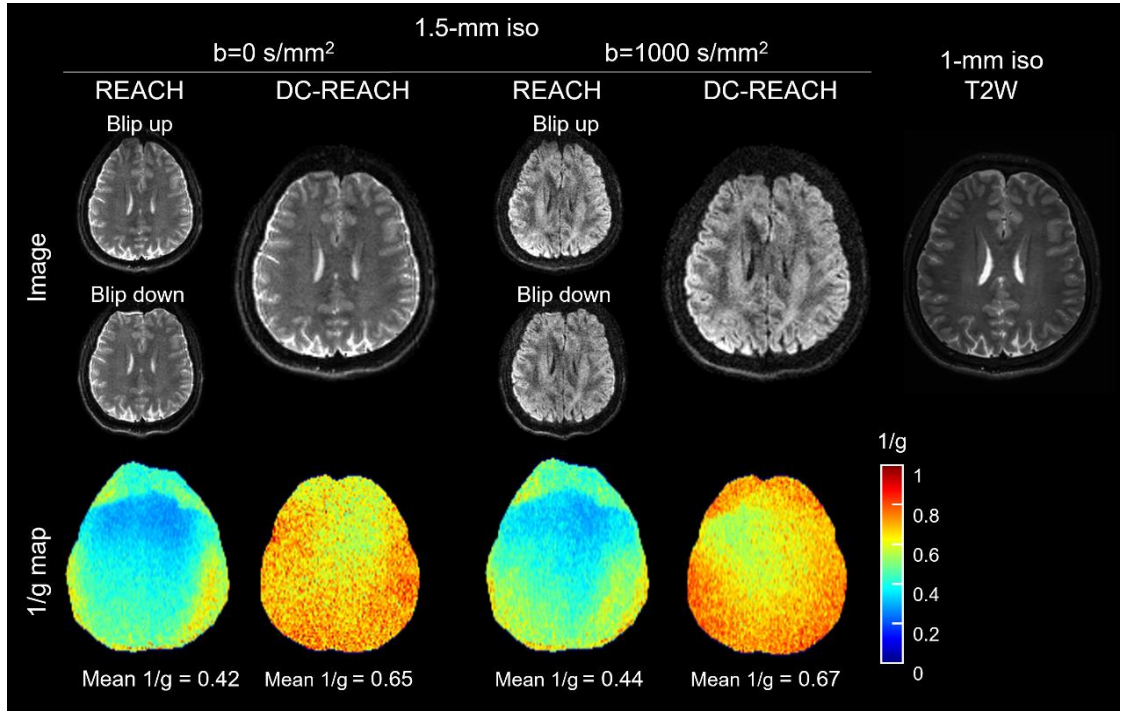

Figure 5: The images of REACH and DC-REACH, which are acquired using blip-up/down blipped-SMSlab with 1.5-mm isotropic resolution. On the top row, the b=0 s/mm² images, b=1000 s/mm² images, and the T2W image are shown from left to right. The corresponding 1/g-factor maps and the mean values of 1/g-factor of one slab are shown at the bottom. For the REACH reconstruction, only the 1/g-factor maps of blip-up images are shown.

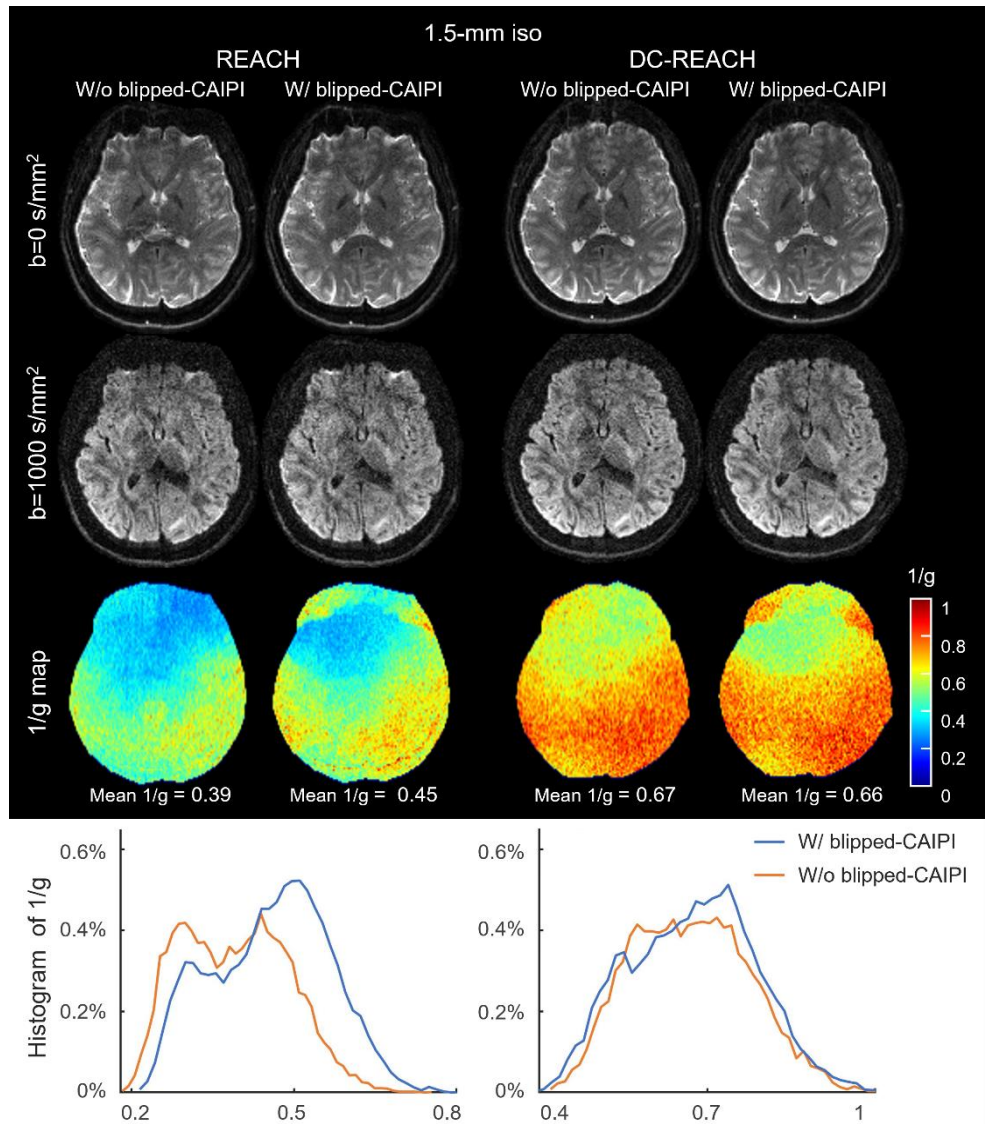

Figure 6: A comparison between SMSlab imaging with and without the blipped-CAIPI gradients. The corresponding trajectories are shown in Supporting Information Figure S6. The blip-down images from REACH alone and the images by DC-REACH from the blip-up/down acquisitions are shown from left to right. The b=0 s/mm$^2$ images, b=1000 s/mm$^2$ images, the corresponding 1/g-factor maps, and the histograms of 1/g-factor are shown from top to bottom. The mean values of 1/g-factor of one slab are marked below the 1/g-factor maps.

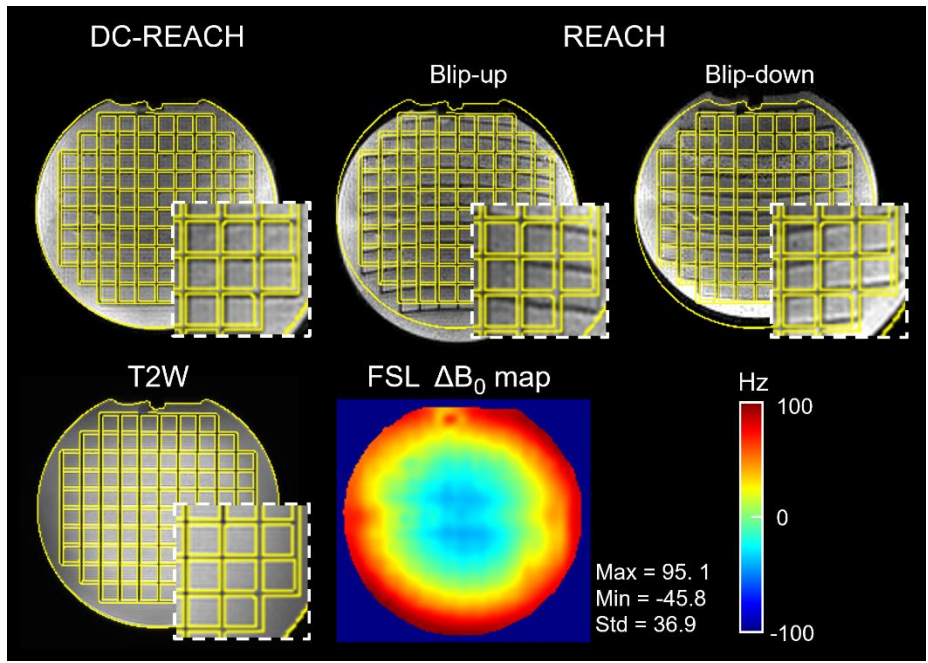

Figure 7: The results of the phantom experiment. On the top row, the b=0 s/mm$^2$ image of DC-REACH, the blip-up image, and the blip-down image of REACH are shown from left to right. The T2W image and the off-resonance map generated by the FSL are shown at the bottom. The edges extracted from the T2W image are superimposed on the b=0 s/mm$^2$ images. The maximum and minimum off-resonance frequencies as well as the standard deviation are marked beside the off-resonance map.

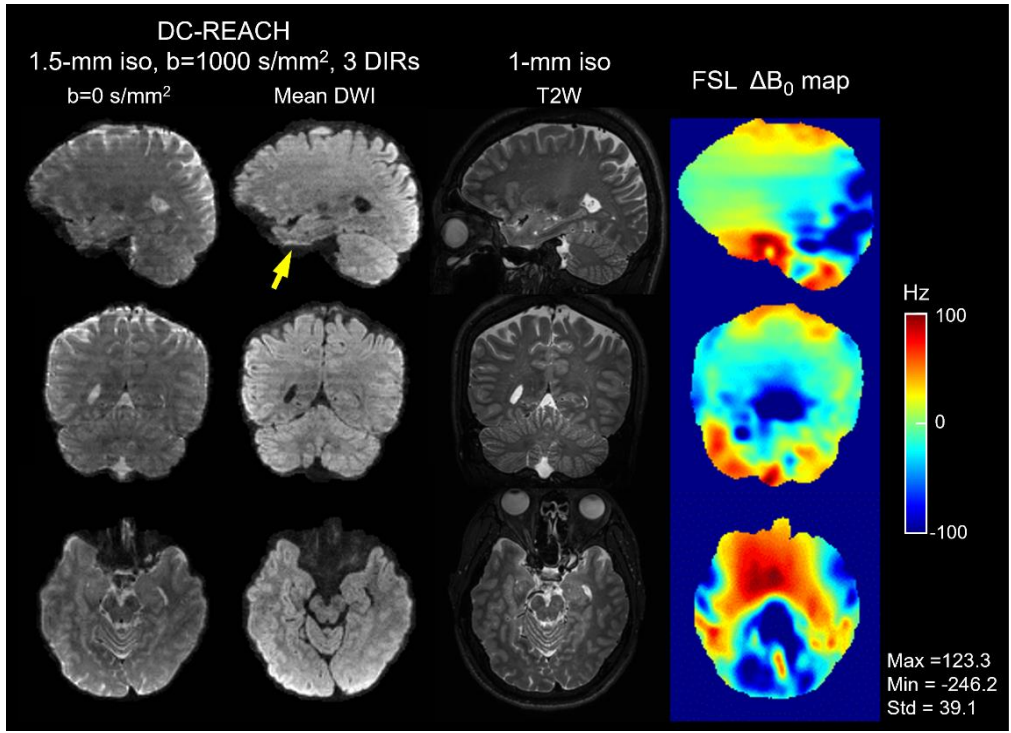

Figure 8: The blipped-SMSlab images with 1.5-mm isotropic resolution and b=1000 s/mm$^2$ reconstructed by DC-REACH. The b=0 s/mm$^2$ images, the mean diffusion-weighted images, the T2W images, and the off-resonance map generated by FSL are shown from left to right. The maximum and minimum off-resonance frequencies as well as the standard deviation are marked beside the maps.

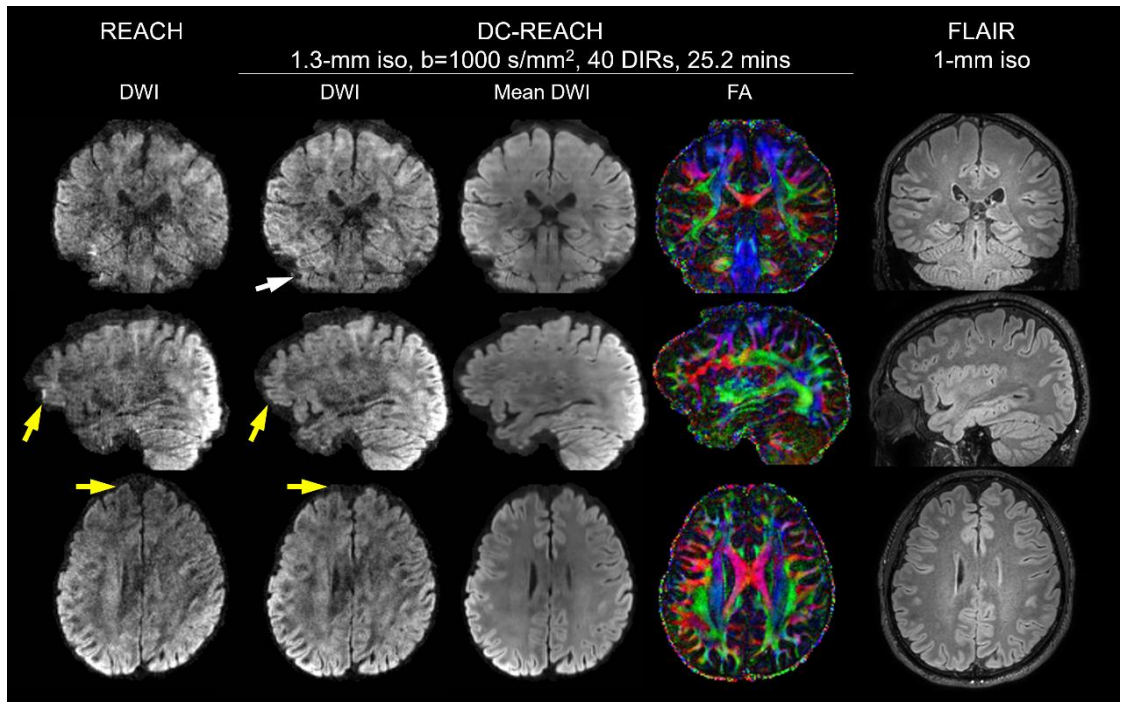

Figure 9: The reconstructed images via DC-REACH for the DTI dataset from blipped-SMSlab with the blip-up/down acquisition. $R_y \times MB=2\times3$, 1.3-mm isotropic resolution, b=1000 s/mm$^2$, and 40 directions were used. The first column shows the distorted diffusion-weighted images reconstructed by REACH. The diffusion-weighted images, the geometric average of 40 directions, and the colored fractional-anisotropy (FA) maps of DC-REACH are shown in columns 2-4. The last column shows the FLAIR images with 1-mm isotropic resolution.

**Supporting Information:**

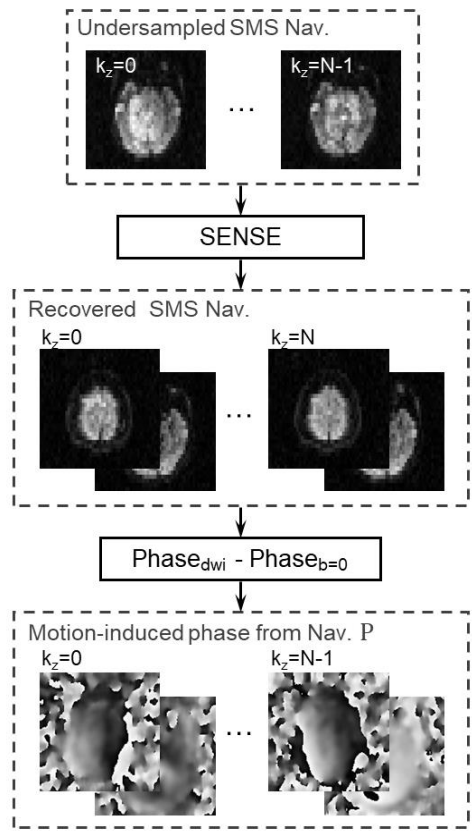

Figure S1: The reconstruction process and the phase extraction of the 2D SMS navigators.

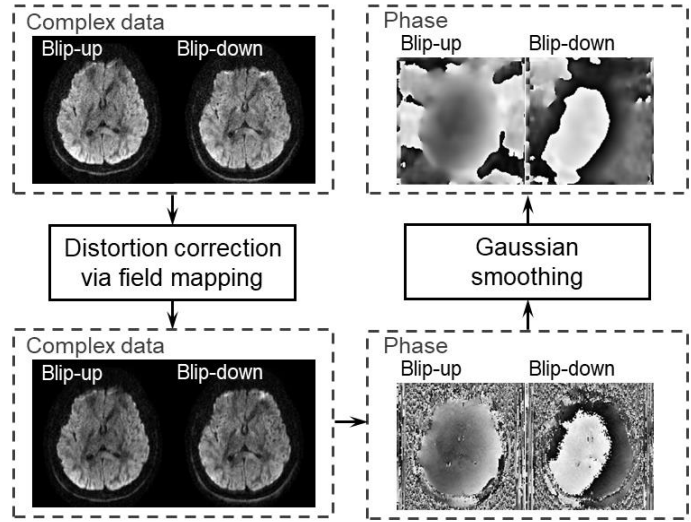

Figure S2: Processing of the phase maps extracted from the blip-up and blip-down images or the recovered navigators in DC-REACH. The field map generated in step 3 of DC-REACH is used for distortion correction via the field mapping method.

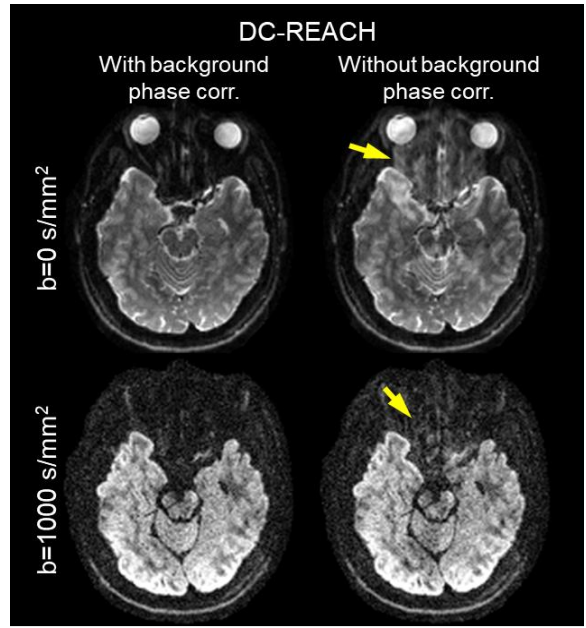

Figure S3: Comparison between the results of DC-REACH with and without background phase correction. Data from the first experiment are shown. Without correcting for the background phase differences between blip-up and blip-down data, inter-slab aliasing artifacts arise in both the b=0 s/mm$^2$ image and the b=1000 s/mm$^2$ image of DC-REACH (yellow arrows).

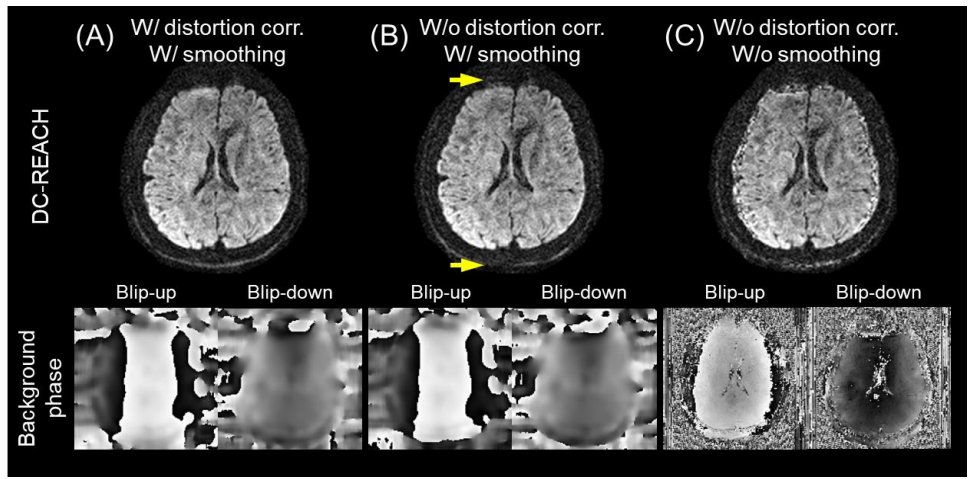

Figure S4: Comparison of the results from different phase maps used in DC-REACH. Data from the first experiment are shown. The reconstruction results are shown at the top, and the corresponding phase maps are shown at the bottom. From left to right, the phase maps are (A) distortion-corrected and smoothed, (B) smoothed but not distortion-corrected, and (C) neither smoothed nor distortion-corrected. Without smoothing or distortion correction, the geometric mismatch between the phase maps of the blip-up/down shots leads to artifacts at the edge of the brain parenchyma (panel C). With the phase maps smoothed but not distortion-corrected (panel B), the results are close to (A) except for some signal loss at the scalp (yellow arrows). The phase maps in (A) for the joint reconstruction provide the best result.

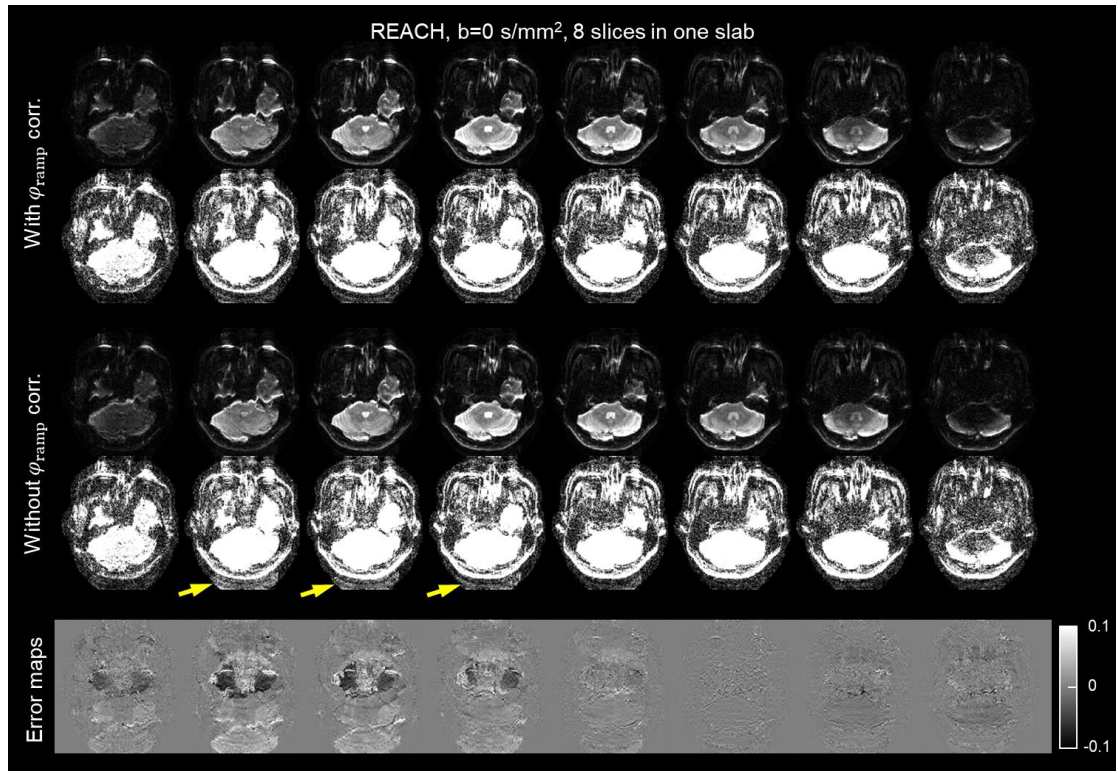

Figure S5: Comparison between blipped-SMSlab images reconstructed with and without $\varphi_{\text{ramp}}$ correction (corr.). The images have 1.5-mm isotropic resolution and only b=0 s/mm$^2$ are shown. Each image is rescaled by a factor of 8. The ghost-like aliasing is pointed out by the yellow arrows. The error maps of the images without $\varphi_{\text{ramp}}$ correction with respect to the images with $\varphi_{\text{ramp}}$ correction are normalized by the maximum amplitude of the images with $\varphi_{\text{ramp}}$ correction.

*SMSlab EPI without blipped-CAIPI and the combination with a blip-up/down acquisition*

The sequence of simultaneous multi-slab (SMSlab) EPI without blipped-CAIPI gradients mainly follows our previous study (28,30). A 3D k-space framework is used. An example with two slabs (slab A and slab B) simultaneously excited is shown in Figures S6B and E. The extra phase offset $\varphi'_{\text{gap}}$ introduced by the existence of the gap between the edge slices of the simultaneously excited slabs ($d_{\text{gap}}$ in Figure 1A) is corrected by RF-based phase modulation. In this way, the simultaneously excited slabs are stitched together and turn out to be a 3D dataset ($k_x$-$k_y$-$k_z$), as shown in Figure S6E. A table of $k_z$ gradients is used to encode the slices in both slabs. It takes $R_{mb} \cdot N_z$ steps of $k_z$ encoding to fully sample the k-space if there are $N_z$ slices in each slab. With multi-band accleration, $N_z$ steps of $k_z$ encoding are applied, which means $k_z$ undersampling with a factor of $R_{mb}$. In this study, only one in-plane shot is acquired for each $k_z$ step. As for the combination with the blip-up/down acquisition, both the blip-up and the blip-down datasets are $R_{mb}$-fold undersampled. A shift along $k_y$ and a shift along $k_z$ are applied for the blip-down dataset. In this way, a CAIPI sampling pattern is formed via the combination of the blip-up and the blip-down datasets.

The reconstruction method is adapted accordingly, which becomes a 3D hybrid-space reconstruction. Because $\varphi_{\text{ramp}}$ correction is not required in SMSlab, only the motion-induced phase variations need to be corrected. The cost function of separate REconstruction with phAse Correction in a Hybrid space (REACH) is

$$I_{\text{REACH}} = \underset{I}{\text{argmin}} \|d - F_u(S \circ P)I\|_2^2. \qquad (1)$$

$F_u$ becomes a 2D FFT operator on y and z dimensions with undersampling along $k_y$. $d$ represents the k-space data that is transformed into the x-$k_y$-$k_z$ space. $S$ represents coil sensitivities. $P$ represents motion-induced inter-shot or inter-$k_z$ phase variations. $I$ is the fully sampled image. $\circ$ means element-wise product. For the distortion-corrected REACH reconstruction (DC-REACH), which jointly reconstructs the blip-up

and blip-down data, the cost function becomes

$$I_{DC} = \underset{I}{\mathrm{argmin}} \|d - (F_u \circ \Theta)(S \circ D \circ P)I\|_2^2. \qquad (2)$$

where $\Theta$ is the field-inhomogeneity-induced phase accrual along $k_y$ dimension which causes image distortions. $\Theta$ is formulated as $\Theta(\Delta f, n_{ky}) = e^{i \cdot \Delta f \cdot (ESP_{eff} \cdot n_{ky})}$, where $n_{ky}$ is the $k_y$ index, and $ESP_{eff}$ is the effective echo spacing. For b=0 data, $\Delta f$ represents the off-resonance field map. For diffusion-weighted data, $\Delta f$ contains both the static off-resonance field and the transient eddy currents field. d is the concatenate data of the blip-up and blip-down shots transformed into the x-$k_y$-$k_z$ space. Some previous results of DC-REACH for SMSlab were submitted as an abstract (52).

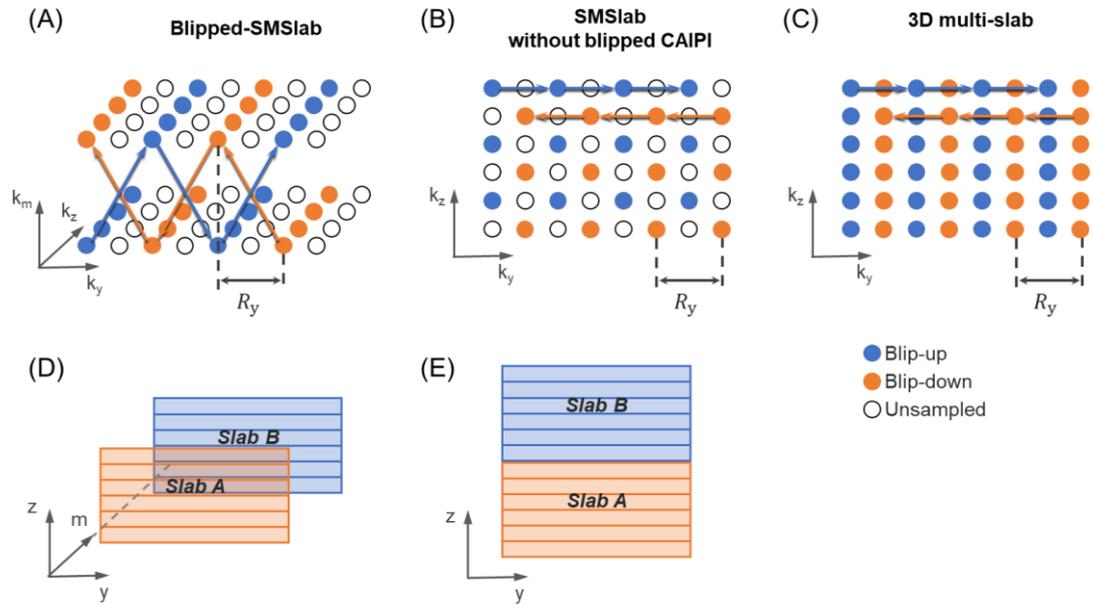

Figure S6: Comparison between blipped-SMSlab (A and D), SMSlab without blipped-CAIPI (B and E), and single-band multi-slab. (A, B, and C) show their trajectories with blip-up/down acquisitions. (D) shows the 4D framework of blipped-SMSlab, in which the simultaneously excited slabs are treated as the fourth logical dimension. (E) shows the 3D framework of SMSlab without blipped-CAIPI, in which the simultaneously excited slabs are stitched together along the $k_z$ dimension. In (D and E), two simultaneously excited slabs are represented by light orange and light blue, respectively. The grids within each slab represent the intra-slab slices.

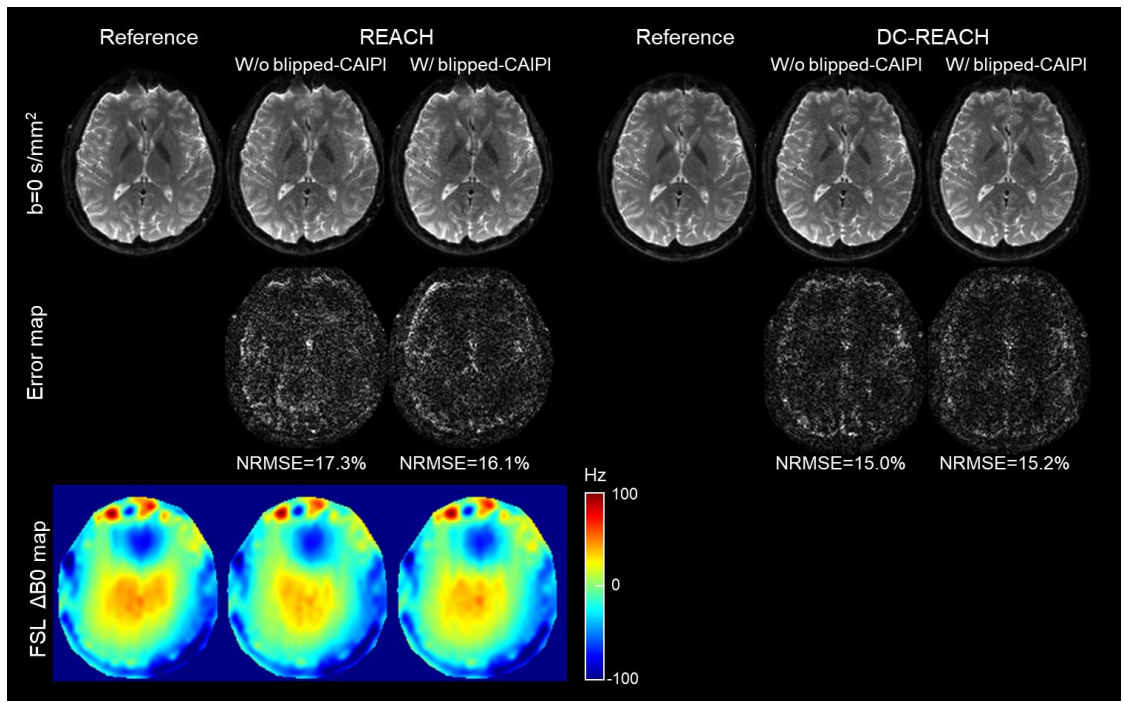

Figure S7: A comparison of reconstruction errors between SMSlab images with and without the blipped-CAIPI gradients. The corresponding trajectories are shown in Supporting Information Figure S6. The images are acquired with 1.5-mm isotropic resolution and b=0 s/mm$^2$. The blip-up images from REACH alone are shown in the upper left. The images reconstructed by DC-REACH from the blip-up/down acquisitions are shown in the upper right. The fully sampled reference image, SMSlab image without blipped-CAIPI, and SMSlab image without blipped-CAIPI (blipped-SMSlab) are shown from left to right. The error maps with respect to the reference are shown in the second row. The off-resonance maps generated by the FSL are shown in the bottom left.

*Adaptation for single-band multi-slab EPI*

REACH and DC-REACH for SMSlab without blipped-CAIPI can be adapted for single-band multi-slab imaging. The trajectory of the blip-up/down acquisition for single-band multi-slab is shown in Figure S6C. For single-band multi-slab imaging, direct undersampling along the $k_z$ direction is not adopted, because the coil sensitivities along the slice direction vary little within a slab on our scanner. Therefore, two in-plane shots are used in each $k_z$ with reversed PE polarities to guarantee the full sampling along $k_z$ for either the blip-up or the blip-down dataset. The signals are encoded in a 3D framework ($k_x$-$k_y$-$k_z$), and the forward model is the same as the model of SMSlab without blipped-CAIPI.

The results are shown in Figure S8. The geometry of the reconstructed images is close to the T2W images. Because no undersampling is used along $k_z$, it takes around 1 minute to acquire one diffusion direction of whole-brain DWI using single-band multi-slab imaging with the blip-up/down acquisition.

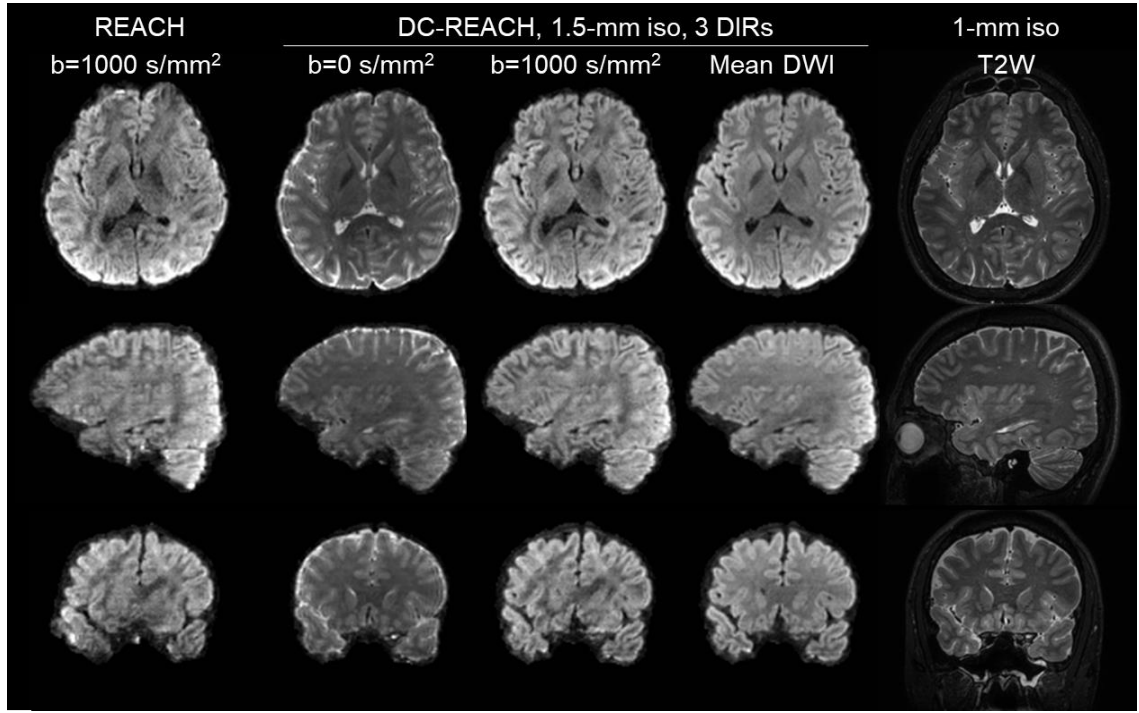

Figure S8: The images of single-band multi-slab EPI with 1.5-mm isotropic resolution. The leftmost column shows the distorted b=1000 s/mm$^2$ images reconstructed by REACH. The middle three columns show the images of DC-REACH. The T2W images are shown on the rightmost.